**Probing Magnetoelastic Coupling and Structural Changes in Magnetoelectric Gallium Ferrite**

Somdutta Mukherjee[1], Ashish Garg[2], Rajeev Gupta[1, 3*]
[1] Department of Physics
[2] Department of Materials Science and Engineering
[3] Materials Science Programme
Indian Institute of Technology Kanpur, Kanpur 208106, India.

**Abstract**
Temperature dependent X-ray diffraction and Raman spectroscopic studies were carried out on the flux grown single crystals of gallium ferrite with Ga:Fe ratio of 0.9:1.1. Site occupancy calculations from the Rietveld refinement of the X-ray data led to the estimated magnetic moment of ~0.60 $\mu_B$ /f.u. which was in good agreement with the experimental data. Combination of these two measurements indicates that there is no structural phase transition in the material between 18 K to 700 K. A detailed line shape analysis of the Raman mode at ~375 cm$^{-1}$ revealed a discontinuity in the peak position data indicating the presence of spin-phonon coupling in gallium ferrite. A correlation of the peak frequency with the magnetization data led to two distinct regions across a temperature ~180 K with appreciable change in the spin-phonon coupling strength from ~ 0.9 cm$^{-1}$ ($T$ < 180 K) to 0.12 cm$^{-1}$ (180 K < $T$ < $T_c$). This abrupt change in the coupling strength at ~180 K strongly suggests an altered spin dynamics across this temperature.

---

[*] Corresponding authors, Email: guptaraj@iitk.ac.in

## 1. Introduction

Gallium ferrite (GaFeO$_3$ or GFO) is a room temperature piezoelectric [1, 2] and a ferri/paramagnet [3]. It possesses a non-centrosymmetric orthorhombic structure with space group *Pc2$_1$n* [4]. Its magnetic transition temperature ($T_c$) is sensitive to the stoichiometry of the material *i.e.* Ga:Fe ratio [5]. Initial interest in GFO arose from a report by Rado [6] showing a large magnetoelectric coupling in stoichiometric GFO ($\alpha_{bc} \approx 1 \times 10^{-11}$ S/m at 77 K). Moreover, in recent years, multiferroics and study of magnetoelectric phenomenon has gained prominence after illustration of this effect in perovskite oxides such as BiFeO$_3$ [7], TbMnO$_3$ [8]. In GFO, recently Ogawa *et al.* [9] demonstrated large magnetoelectric coupling as shown by magnetization induced second harmonic generation and large angle Kerr rotation of the second harmonic light in the ferrimagnetic state. In another study, Jung *et al.* [10] reported optical absorption in the presence of a magnetic field of 500 Oe suggesting a large optical magnetoelectric effect below $T_c$. Moreover, in materials exhibiting large magnetoelectric coupling [11, 12], one may expect structural and magnetic degrees of freedom behaving in tandem. However, while GFO's magnetoelectric characteristics are reasonably well demonstrated, its magnetoelastic behavior demands studies on high quality samples to elucidate structure-property correlations in GFO.

A recent structural study [10] on single crystal GFO using neutron and synchrotron scattering has reported absence of any structural change across the magnetic transition. In addition, significant distortion of the FeO$_6$ octahedra in the unit-cell resulted in the shift of Fe ions from the center of FeO$_6$ octahedra, leading to a spontaneous polarization along the crystallographic *b*-direction [5]. In the context of structure related studies of GFO, we note that many such studies have been conducted using tools which yield information at a rather macroscopic scale. It would, therefore, be interesting to examine the structure of GFO at nm-length scales using techniques such as Raman spectroscopy. Further, to the best our knowledge, there is no report on structural study of GFO beyond room temperature. Such a study, both at macroscopic and microscopic length scales, could shed light on the possible temperature driven structural phase transformation imparting higher symmetry to the system.

Raman spectroscopy has been used extensively to probe intricate structural details of various multiferroic materials such as structural distortion [13], spin dynamics [14] and any kind of coupling between structure and magnetic degrees of freedom [15, 16]. Spin-phonon coupling has been demonstrated using Raman spectroscopy in a variety of materials: in rare earth manganites [17] showing large phonon softening across magnetic $T_c$, in TbMnO$_3$ [15] and in BiFeO$_3$ [18]. In rare-earth manganites, it was also shown that the manganites with smaller rare earth ion show negligible spin lattice coupling [17]. All these reports conclusively prove the utility of Raman spectroscopy to investigate multiferroic materials. In this manuscript, we report a detailed x-ray diffraction and Raman study on flux-grown GFO single crystals to investigate the structural evolution and role of low energy excitations across different transitions as a function of temperature in the temperature range 18 K to 700 K.

## 2. Experimental Details

Single crystals of GFO were synthesized by flux growth process using β-Ga$_2$O$_3$, α-Fe$_2$O$_3$ as constituents and Bi$_2$O$_3$ and B$_2$O$_3$ as flux in weight ratio 1: 1: 5.4:0.6 using the methodology similar to that reported in Ref. [1]. Dark brown crystals with dimensions ~ 2 × 4 × 1 mm were extracted. Compositional analysis of the samples was carried out using energy dispersive spectroscopy (EDS) using Oxford EDS Spectrometer in a Zeiss scanning electron microscope. Average EDS data indicates that crystals are chemically homogeneous with the Ga to Fe ratio of ~ 0.93: 1.11 vis-à-vis initial mixing stoichiometry of 0.93: 1.09 (Ga:Fe). Therefore, we specify our samples as $Ga_{2-x}Fe_xO_3$ with $1.09 \leq x \leq 1.11$. Temperature dependent powder X-Ray Diffraction (XRD) of the crushed single crystals was carried out using a high resolution Philips X'Pert PRO MRD diffractometer with an angular resolution of 0.0001°. Micro-Raman study was carried out in back scattering geometry using a high resolution (~ 0.6 $cm^{-1}$/pixel) system equipped with $Ar^+$ 514.5 nm laser as an excitation source and a liquid nitrogen cooled CCD detector. The sample temperature was varied between 18 K - 450K using a closed cycle cryostat for low temperature region while a hot-cold Linkam THMS600 stage was used for high temperature region. Magnetization data as a function of temperature was acquired using a Quantum Design SQUID magnetometer.

## 3. Results and Discussions
### 3.1 X-ray Diffraction Analysis

XRD data of the crushed single crystals were acquired from 300 to 700 K at 50 K interval and within the 2θ region of 26°-120°. Fig. 1 (a) shows representative XRD plots of the data taken at room temperature and 700 K. The spectra were indexed to GFO ICDD data card no. 76-1005 and the peak match suggests single phase formation with no signature of intermediate or secondary phase(s). Same is true for all other temperatures whose plots are not shown here. Further, to understand any subtle changes in the structure as well as to quantify the occupancy of lattice sites, we carried out Rietveld refinement of XRD data using FULLPROF 2000 package [19]. We used orthorhombic $Pc2_1n$ ($Pna2_1$: notation used in international table of crystallography) symmetry of GFO for refinement at all temperatures along with the site occupancies determined at room temperature. To avoid ambiguities in the structural refinement, we excluded the 2θ region between 30.2°-32.45° consisting of a major peak of the high temperature dome material (PEEK). The refinement of the XRD data, also shown in Fig. 1(a), did not suggest any major change in the peak profiles or relative intensities, ruling out any anomalous structural change or distortion in GFO over the temperature range of 300-700 K. The resolution of our measurements is better than 0.003Å for all the *d*-values, well within the instrumental resolution. Further we analyzed the refined XRD data to understand the effect of temperature on the unit-cell parameters, bond lengths and polyhedral distortion.

From the refinement, we obtained the atomic positions of the constituent ions and refinement parameters (not shown here). Further, we estimated the room temperature lattice parameters of GFO from the refined data: *a* = 8.744 Å, *b* = 9.395 Å, *c* = 5.079 Å and unit-cell volume, $V_{cell}$, of 417.26 $Å^3$. These values are in excellent agreement with those reported previously [4, 5]. Similarly, at 700 K the unit cell parameters and volume, calculated for the refined structure are *a* = 8.769 Å, *b* = 9.422 Å, *c* = 5.097 Å and $V_{cell}$ = 421.12 $Å^3$, respectively. We investigated the variation in the lattice parameters and unit cell volume at temperatures from 300-700K and a plot of these parameters versus temperature is shown in Fig. 1(b). As can be seen from the plot, the lattice parameters vary linearly with temperature. From the linear fitting of above data, the coefficients of thermal expansion ($\alpha$, K$^{-1}$) of GFO along three principal axes ($a$, $b$ and $c$) were calculated to be 7.2×10$^{-6}$, 7.5×10$^{-6}$ and 8.8×10$^{-6}$, respectively. Negligible difference among the three values suggests that GFO is a thermally isotropic material. From the slope of unit cell volume vs. temperature plot, the thermal coefficient for volume expansion ($\gamma$) was calculated to be 2.37×10$^{-5}$ K$^{-1}$ which follows the relation $\gamma = 3\alpha$, true for an isotropic material. Since, low temperature XRD experiments could not be carried out; we estimated the lattice parameters at low temperature by extrapolation. The estimated lattice parameters at 4 K i.e. $a \sim 8.725$ Å, $b \sim 9.374$ Å and $c \sim 5.066$ Å, are in excellent agreement with those reported by Arima *et al.* [5], ($a \sim 8.71932(13)$ Å, $b \sim 9.36838(15)$ Å $c \sim 5.06723(8)$ Å). The fact that extrapolated lattice parameters agree with the reported low temperature data, it can be concluded that there is no structural distortion or lattice parameter anomaly in GFO down to 4K.

Further, the refined structural data were used to calculate the bond lengths using an approach used by Momma and Izumi [20]. For clarity, a simulated crystal structure of GFO has been shown in Fig. 1(c) where positions of Fe and Ga ions are marked: Ga1 is tetrahedrally coordinated while Ga2, Fe1 and Fe2 are octahedrally coordinated by oxygen atoms. Fig. 1(d) shows four different cation-oxygen polyhedra at room temperature. It was found that Ga1–O tetrahedron shares its corners with surrounding cation octahedra which share their edges among neighboring octahedra. The estimated average bond lengths for Ga1-O, Ga2-O, Fe1-O and Fe2-O ions at room temperature are 1.860 Å, 2.023 Å, 2.033 Å and 2.039 Å, respectively. It was observed that upon increasing the temperature from 300 K to 700 K, Ga1-O, Ga2-O and Fe1-O bonds stretched by 0.3 %, 0.23 % and 1.25 %, respectively while Fe2-O bond contracted by 0.21 %. This translates to 0.54% reduction in the volume of Fe2-O octahedron, and an increase of 0.75%, 0.45% and 2.9 % in Ga1-O tetrahedron, Ga2-O and Fe1-O octahedron volume, respectively. Further, the changes in the bond lengths result in polyhedral distortion which is quantified by calculating the distortion index ($\Delta$) as defined by Baur [21];

$$\Delta = \frac{1}{n} \sum_{i=1}^{n} \frac{(l_i - l_{avg.})}{l_{avg.}} \qquad (1)$$

where $l_i$ = distance from the central atom to the $i^{th}$ coordinating atom, $n$ is the number of bonds and $l_{avg.}$ = average bond length. It is observed that, at room temperature, Ga1-O tetrahedron is least distorted ($\Delta = 0.0065$) among all the polyhedra and remains close to its ideal shape. While, both Fe1-O and Fe2-O octahedra are significantly distorted with $\Delta = 0.0650$ and $0.0760$, respectively and Ga2-O octahedron is comparatively less distorted ($\Delta = 0.0182$). Upon increasing the temperature to 700 K, the distortion indices of the above polyhedra change to 0.0051, 0.0806, 0.0701 and 0.0260, respectively. This shows that with increasing temperature Ga1-O tetrahedron tends to move toward the shape of a regular tetrahedron ($\Delta = 0.0$), the shape of Fe2-O octahedron remains almost identical to its RT structure and Ga2-O and Fe1-O octahedra are further distorted. Such details on the structural distortion of GFO can be crucial to understand the temperature dependence of microscopic polarization in GFO. Previous first principle studies have shown important correlation between the structural distortion and polarization in GFO [22].

Structural refinement of the XRD data also gives an estimate of the cation occupancies which can prove important in explaining the observed magnetic behavior as cation site disordering can lead to significant changes in the magnetic properties [3, 5]. Site disordering in GFO is expected due to the fact that, in addition to being isovalent, the ionic radii of Fe and Ga are quite close to each other (0.645 and 0.62 Å, respectively). From the room temperature data, we refined the cation occupancies keeping anion occupancies fixed at 1.0. The occupancies of iron at Ga1, Ga2, Fe1 and Fe2 sites are found to be 0.14, 0.32, 0.83 and 0.89, respectively. From the partial occupancies, we calculated Ga to Fe ratio in GFO as ~ 0.93:1.11 which is in very good agreement with the EDS data, as mentioned in section 2. In the following paragraph, we discuss the correlation between cation site occupancies and the observed magnetic behavior of the GFO samples.

If GFO were to behave as a perfect antiferromagnet, the Fe spins on two octahedral sites would be antiparallel resulting in a net zero magnetic moment [5, 23]. In contrast, experimental studies show a large magnetic moment in GFO below $T_c$ [5, 24]. It has been suggested that this observed magnetic moment at low temperature can be attributed to the cation site disorder, *i.e.*, presence of octahedral Fe ions predominantly, on the octahedral Ga sites [5]. Our temperature dependent magnetization measurement (not shown here) showed a ferri- to para-magnetic transition at ~ 290 K and also yielded a magnetic moment per Fe site of ~ 0.67 $\mu_B$ at 4 K. Previous neutron diffraction study [5] shows Fe at Fe2 and Ga2 sites are ferromagnetically coupled. However, neutron diffraction data did not comment about the spin configuration of Fe at Ga1 site. Assuming ferromagnetic coupling of Fe at Fe1 and Ga1 sites and high spin moment of Fe [3] and, using the cation partial occupancies from our Rietveld data, we estimated net magnetic moment of GFO to be ~ 0.60 $\mu_B$/ f.u. which is in excellent agreement with our experimental observation. On the contrary, if we assume Fe at Ga1 and Fe1 are aniferromagnetically coupled, the estimated moment is ~ 1.3 $\mu_B$/ f.u. which is very large in comparison to our experimental data. Thus, we conclude that Fe at Ga1 site is ferromagnetically aligned with respect to Fe at Fe1 site and antiferromagnetically aligned to Fe at Fe2 (and also Fe at Ga2) site. We also calculated net magnetic moment using partial occupancies and magnetic moments determined by Arima *et al* [5]. These results are tabulated in Table 1. Based on our assumption that Fe at Fe1 and Ga1 sites are ferromagnetically coupled and using the site occupancy data in stoichiometric GFO as reported by Arima *et al.* [5], we estimated the net magnetic moment of stoichiometric GaFeO$_3$, as ~ 0.55 $\mu_B$/ f.u. which is close to their experimental observation (~ 0.65 $\mu_B$/ f.u.). Here we assumed that the moment of Fe at Ga1 site is same as the moment of Fe at Fe1 site. Finally, we calculated the net magnetic moment of GFO (0.86 $\mu_B$/ f.u.) using partial occupancies of our Rietveld data and the magnetic moments from Arima *et al.* [5]. The observed difference between the calculated and experimental moments might be attributed to the difference in the magnetic moments of Fe at cation sites (with respect to Arima *et al.* [5]) due to nonstoichiometry of our sample. From this discussion, we conclude that the magnetism in GFO is highly sensitive to the cation site occupancies. Thus it is possible to tune the magnetic behavior of GFO by careful compositional tailoring.

## 3.2 Temperature dependent Raman Spectroscopy

In a recent theoretical work Fennie *et al.* [25], predicted that combination of strain, spin-phonon coupling and optical modes can play an important role in simultaneously stabilizing both ferroelectric and ferromagnetic phases. The veracity of Fennie *et al.*'s work was demonstrated in a recent report by Lee *et al.* [11] in which tuning of bi-axial strain in EuTiO$_3$ thin films led to a large spin-lattice coupling resulting in simultaneous ferroelectricity and ferromagnetism. Further, since Raman spectroscopy can probe lattice excitations *i.e.* phonons, and magnetic excitations *i.e.* magnons as well as their interactions, it is an ideal tool to investigate the spin-lattice coupling in materials. Spin correlation among the nearest neighbors *i.e.* <$S_i.S_j$> can be used to relate the behavior of the phonons to determine the spin-lattice coupling strength [26].

In GFO, while the neutron studies made by Arima *et al.* did not make any mention of the first-order spin-lattice coupling [5], a recent Raman study on polycrystalline bulk stoichiometric GFO [27] speculated the existence of spin-lattice coupling based on the discontinuity in the line width of one of the phonons at 200 K. However, poor quality of Raman spectra, possibly due to polycrystalline nature of the samples, does not exude the confidence to draw these conclusions. In contrast, our temperature dependent Raman measurements have been carried out on high quality single crystals of GFO to investigate temperature dependent behavior of Raman modes from the perspective of understanding the role of phonons across the phase transitions in this material.

Group theoretical methods predict that there are total of 120 normal modes (117 Raman and 97 IR active) of vibration in GFO considering 8 f.u. in a primitive cell and *Pna2$_1$* ($C_{2v}^9$) space group [27]. Since above space group does not contain an inversion center, the IR active modes are simultaneously Raman active and these modes are non-degenerate [28]. In the present study, we acquired Raman spectra of single crystals and observe 31 Raman active modes in the spectral range of 90-900 cm$^{-1}$ at room temperature. It is likely that while some modes are below our detection limit on the lower wave number side, orientation of the crystal may also restrict the observation of some other modes due to selection rules.

As can be seen from Fig. 2(a), 31 distinct Raman peaks occur at 99, 106, 118, 121, 129, 138, 149, 154, 173, 199, 211, 219, 240, 257, 270, 303, 350, 359, 370, 394, 438, 465, 521, 575, 604, 655, 674 688, 723, 743 and 759 cm$^{-1}$. In contrast, the report by Sharma *et al.* [27] on polycrystalline GFO samples showed a rather broad spectra with fewer number of Raman modes. The representative spectra measured on our samples at various temperatures between 18-450 K are shown in Fig. 2(a). A first glance suggests that the Raman modes harden with lowering of temperature accompanied by narrowing of the peak widths. Moreover, upon cooling across the magnetic transition ($T_c$ ~ 290 K), the number of Raman peaks remains same suggesting absence of a structural change near $T_c$, consistent with temperature dependent Neutron diffraction studies reported earlier [5]. For further analysis of the Raman data to examine any subtle changes in the structure, we divided the entire spectrum into two ranges *i.e.* 275-550 cm$^{-1}$ and 620-820 cm$^{-1}$. The spectra in these two ranges were deconvoluted into a sum of nine and six Lorentzians, labeled as M1, M2, …., M9 and X1, X2,….,X6 respectively, as shown in Fig. 2 (b) and (c) along with the fitted curves. The extracted line shape parameters characterizing a Lorentzian function *i.e.* peak positions and line widths for modes M1-M9 are plotted as a function of temperature in Fig. 3. The figure depicts that the peak positions shift towards lower frequencies and the line widths or FWHM increase with increasing temperature. Similar behavior also holds true for X1-X6 modes. Such a behavior is an expected outcome of lattice expansion and increase in the phonon population as a result of increasing temperature.

This hardening of phonons with decreasing temperature can be described by the following relation [29],

$$\omega_j(T) = \omega_j(0) + (\Delta\omega_j)_{qh}(T) + (\Delta\omega_j)_{anh}(T) + (\Delta\omega_j)_{el-ph}(T) + (\Delta\omega_j)_{sp-ph}(T) \qquad (2)$$

where $\omega_j(T)$ and $\omega_j(0)$ are the phonon frequencies of $j$ th mode at any temperature T and at 0 K, $(\Delta\omega_j)_{qh}$, $(\Delta\omega_j)_{anh}$, $(\Delta\omega_j)_{el-ph}$ and $(\Delta\omega_j)_{sp-ph}$ are the changes in phonon frequencies, respectively due to change in the lattice parameters of the unit cell (quasi-harmonic effect); intrinsic anharmonic interactions; electron-phonon coupling and spin-phonon coupling in magnetic materials caused by the modulation of exchange integral by lattice vibrations [29].

The quasi-harmonic effect on the vibrational mode due to change in the unit cell volume triggered by thermal effects can be approximated by Grüneisen law relating the change in the frequency to the change in the lattice volume [29] i.e. $\left(\Delta\omega_j/\omega_j\right)_{qh} = -\gamma_j \left(\Delta V/V\right)$ ($\gamma_j$: Grüneisen parameter for the normal mode $j$ and $\Delta V/V$: the fractional unit cell volume change due to thermal expansion). Since, XRD analysis shows minute change (<1%) in the unit cell volume of GFO over the temperature range of 18 K to 450 K, therefore, it is expected that the contribution of quasiharmonic effect on the mode frequency is negligible.

Next, we examine the contribution due to intrinsic anharmonic interactions i.e. $(\Delta\omega_j)_{anh}$. Considering only the contribution from cubic and quartic anharmonic processes and further assuming that each phonon decays into two lower energy phonons of equal energy i.e. a phonon with frequency $\omega$ decays into two (three) phonons of frequency $\omega/2$ ($\omega/3$) for the cubic (quartic) anharmonic process. Therefore, the temperature dependence of frequency of $j^{th}$ mode can be represented by the following relation (eq. (3)) [30]

$$\omega_j(T) = \omega_j(0) - A_j\left[1 + \frac{2}{e^{\hbar\omega_j(0)/2k_BT} - 1}\right] - B_j\left[1 + \frac{3}{e^{\hbar\omega_j(0)/3k_BT} - 1} + \frac{3}{(e^{\hbar\omega_j(0)/3k_BT} - 1)^2}\right] \qquad (3)$$

where, $\omega_j(0)$ is 0 K frequency of the mode in harmonic approximation while $A_j$ and $B_j$ are anharmonicity coefficients, giving the strength of cubic and quartic anharmonic processes, respectively.

Similarly, the line width of a Raman mode has following temperature dependence:

$$\Gamma_j(T) = \Gamma_j(0) + C_j\left[1 + \frac{2}{e^{\hbar\omega_j(0)/2k_BT} - 1}\right] + D_j\left[1 + \frac{3}{e^{\hbar\omega_j(0)/3k_BT} - 1} + \frac{3}{(e^{\hbar\omega_j(0)/3k_BT} - 1)^2}\right] \qquad (4)$$

where, $\Gamma_j(0)$ is intrinsic broadening of $j^{th}$ mode arising from factors other than phonon decay such as presence of structural defects while $C_j$ and $D_j$ are parameters for cubic and quartic anharmonic processes, respectively.

Equations (3) and (4) have been used to fit the peak frequency and line width data for modes M1-M9 (shown in Fig. 3) and modes X1-X6 (not shown here). It was observed that temperature dependence of line shape parameters of modes M1, M2, M3, M4, M6 (see Fig. 3), fits reasonably well to eq. (3) and (4). However, the frequency shifts of modes M5, M7, M8 and M9 cannot be adequately described by eq. (3) using one set of fitting parameters. Hence, the parameters $\omega(0)$, $A$ and $B$ for these modes were optimized by fitting the experimental data above $T_c$. Below $T_c$, the most pronounced deviation from the fit is found for modes M5 and M9, with M5 softening and M9 hardening. Fig. 3 also shows the plot of variation of line width (right axis) of the above nine modes (M1-M9) as a function of temperature along with the fits using eq. (4). Here, the modes M2, M3, M4, M6 and M8 in Fig. 3 particularly fit well to eq. (4) throughout the entire temperature range. As far as modes M1, M5, M7, M9 are concerned, the line widths of these modes follow the anharmonic interaction model reasonably well below $T_c$. However, above $T_c$, line widths of these modes become nearly temperature independent resulting in significant deviation from the fit to eq. (4). These changes in the line widths for some of the modes below $T_c$ indicate towards a spin-phonon interaction in GFO. The apparent temperature independent behavior of line widths of a few modes can be argued as a consequence of the competition between decrease in the line widths due to absence of magnon-phonon interaction above $T_c$ and increase in the line widths with increased site disordering with increasing temperature. Since, GFO is an insulator within the studied temperature range, electron-phonon interaction is unlikely to be temperature dependent and hence its contribution to change in line width with temperature can be neglected [29].

Finally, to quantify the spin-phonon coupling strength as given by the last term in eq. (2), we utilize the formalism proposed by Granado *et al.* [29] suggesting a mechanism for phonon renormalization due to spin-phonon interactions. The spin-phonon coupling strength for a given mode can be estimated by relating the change in the peak position from the conventional anharmonic behavior below $T_c$ to the nearest neighbor spin-spin correlation function as given by $\Delta\omega_{spin-phonon} \approx \eta \langle \vec{S}_i \cdot \vec{S}_j \rangle$. Here, $\langle \vec{S}_i \cdot \vec{S}_j \rangle = 2\{M(T)/M_s\}^2$; M(T) is the magnetization of GFO per Fe site below $T_c$, $M_s$ is the saturation magnetization and $\eta$ represents the spin-phonon coupling strength. The factor of two on the right arises due to two nearest neighbors in the ferromagnetic *ac*-plane for each type of Fe ion in the unit-cell. Since, mode M5 at 374 cm$^{-1}$ exhibits the largest deviation from the conventional anharmonic behavior below $T_c$, this should correspond to the possibly largest value of $\eta$. As per above relation, a plot between $\omega(T)$ vs. $\{M(T)/M_s\}^2$, as shown in Fig. 4, is used to estimate the value of $\eta$ (½ of the slope, $m$) for GFO. The figure shows two distinct regions across a temperature ~180 K, defined as $T_f$, as depicted by the sharp change of slope: a region on the low temperature side below $T_f$ with relatively large value of $\eta \sim 0.9$ cm$^{-1}$ and another region above $T_f$ but below $T_c$ with a much smaller value of $\eta \sim 0.12$ cm$^{-1}$. Such an abrupt change in $\eta$ across $T_f$ is a strong indication of change in the spin dynamics in GFO. Moreover, these values of $\eta$ obtained for GFO are comparable to earlier estimates reported on other systems using Raman scattering. For instance, on antiferromagnetic rutile structured $MnF_2$ and $FeF_2$, Lockwood *et al.* [26] showed the spin phonon coupling strength for different modes in the range from 1.3 cm$^{-1}$ to 0.4 cm$^{-1}$ while Gupta *et al.* [31] found a very large spin phonon coupling strength (~ 5.2 cm$^{-1}$) for $Sr_4Ru_3O_{10}$. Therefore, our measurements suggest that the spin-phonon coupling of GFO is rather weak immediately below $T_c$ until $T_f$ before increasing substantially below $T_f$. It would be interesting to further examine this phenomenon across $T_f$ by carrying out *ac* magnetic measurements as a function of temperature and frequency to understand the spin dynamics and to probe deeper into the nature of this transition.

**Conclusions**

In conclusion, temperature dependent X-ray and Raman studies of GFO single crystals ruled out any structural transition between 18 K and 700 K. Rietveld refinement of the XRD data showed a thermally isotropic nature of the material. Calculated magnetic moment based on the cation occupancies, determined from Rietveld refinement, matched very well with the experimentally measured values. The variation of the peak position for the Raman mode at ~ 374 cm$^{-1}$ with temperature suggested spin-phonon interactions in the material with a coupling strength of ~ 0.9 cm$^{-1}$ below ~180 K. The abrupt change in the slope of phonon frequency versus the square of normalized magnetization at ~180 K indicates the change in the nature of spin-lattice interactions across this temperature.

**Acknowledgements**
Authors acknowledge the financial support from Council of Scientific and Industrial Research (CSIR) and Department of Science and Technology (DST), Govt. of India.

**References:**
1. Remeika J P 1960 *J. Appl. Phys.* **31** S263.
2. White D L 1960 *Bull. Am. Phys. Soc.* **5** 189.
3. Frankel R B, N A Blum, Foner S, Freeman A J and Schieber M 1965 *Phys. Rev. Lett.* **15** 958.
4. Abrahams S C, Reddy J M and Bernstein J L 1965 *J. Chem. Phys.* **42** 3957.
5. Arima, T *et al Phys. Rev. B* 2004 **70** 064426.
6. Rado G T 1964 *Phys. Rev. Lett.* **13** 335.
7. Wojdeł J C and Íñiguez J 2009 *Phys. Rev. Lett.* **103** 267205.
8. Kimura T, Goto T, Shintani H, Ishizaka K, Arima T and Tokura Y 2003 *Nature* **426** 55.
9. Ogawa Y, Kaneko Y, He J P, Yu X Z, Arima T and Tokura Y 2004 *Phys. Rev. Lett.* **92** 047401.
10. Jung J H, Matsubara M, Arima T, He J P, Kaneko Y and Tokura Y 2004 *Phys. Rev. Lett.* **93** 037403.
11. Lee, J H *et al* 2010 *Nature* **466** 954.
12. Oh Y S *et al* 2011 *Phys. Rev. B* **83** 060405.
13. Singh M K, Ryu S and Jang H M 2005 *Phys. Rev. B* **72** 132101.
14. Suzuki N and Kamimura H 1973 *J. Phys. Soc. Japan* **35** 985.
15. Rovillain P, Cazayous M, Gallais Y, Sacuto A and Measson M-A 2010 *Phys. Rev. B* **81** 054428.
16. Ferreira W S *et al* 2009 *Phys. Rev. B* **79** 054303.
17. Laverdière *et al* 2006 *Phys. Rev. B* **73** 214301.
18. Singh M K, Katiyar R S, and Scott J F 2008 Xiv:0712.4040v2 [cond-mat.mtrl-sci].
19. Rodríguez-Carvajal J 1993 *Physica B: Condensed Matter* **192** 55.
20. Momma K and Izumi F 2008 *J. Appl. Crystallogr.* **41** 653.
21. Baur W H 1974 *Acta Crystallogr.* B **30** 1195.
22. Roy A, Mukherjee S, Gupta R, Auluck S, Prasad R and Garg A 2011 *J. Phys.: Condens. Matter* 23 325902.
23. Han M J, Ozaki T and Yu J, 2007 *Phys. Rev. B* **75** 060404.
24. Nowlin C H and Jones R V 1963 *J. Appl. Phys.* **34** 1262.
25. Fennie C J and Rabe K M 2006 *Phys. Rev. Lett.* **97** 267602.
26. Lockwood D J and Cottam M G 1988 *J. Appl. Phys.* **64** 5876.
27. Sharma K, Reddy R, Kothari D, Gupta A, Banerjee A and Sathe V G 2010 J. *Phys.: Condens. Matter* **22** 146005.
28. Xia H R *et al* 2004 *J. Raman Spectroscopy* **35** 148.
29. Granado E *et al* 1999 *Phys. Rev. B* **60** 11879.
30. Balkanski M, Wallis R F and Haro E 1983 *Phys. Rev. B* **28** 1928.
31. Gupta R, Kim M, Barath H, Cooper S L and Cao G 2006 *Phys. Rev. Lett.* **96** 067004.

List of Tables

Table 1 – Estimation of magnetic moments (μ) in GFO using site occupancies (occ.) calculated from the XRD data and comparison with those calculated using Arima *et al.*'s [5] data. (u.c.: unit-cell and f.u.: formula unit)

| Site (Occ.*) | Fe Occ.** | Fe Occ.† | μ ($Fe^{3+}$) ($μ_B$)** | μ ($Fe^{3+}$) ($μ_B$)† | Total μ/u.c. ($μ_B$) | | |
|---|---|---|---|---|---|---|---|
| | | | | | ** | $ | † |
| Ga1 (0.86) | 0.14 | 0.18 | -5 | -3.9 | -2.8 | -2.18 | -2.81 |
| Ga2 (0.68) | 0.32 | 0.35 | 5 | 4.7 | 6.4 | 6.2 | 6.58 |
| Fe1 (0.83) | 0.83 | 0.77 | -5 | -3.9 | -16.6 | -12.95 | -12.01 |
| Fe2 (0.89) | 0.89 | 0.70 | 5 | 4.5 | 17.8 | 16.02 | 12.60 |
| | | | | | 4.8 | 7.09 | 4.36 |
| | | | | | 0.6 (/f.u.) | 0.89 (/f.u.) | 0.54 (/f.u.) |

*Cationic site occupancies obtained after Rietveld refinement
** Present calculation assuming high spin $Fe^{3+}$ moment of 5$μ_B$
† Calculated Fe site occupancies and magnetic moment using Neutron diffraction data by Arima *et al.* [5]
$ Calculation assuming Fe site occupancies from the present study and $Fe^{3+}$ magnetic moments by Arima *et al.* [5]

**List of Figures**

Fig. 1. (Color online) (a) Rietveld refined XRD patterns of crushed GFO single crystals at 300 K and 700 K, (b) lattice parameters and unit cell volume obtained from Rietveld refinement plotted as a function of temperature, (c) simulated crystal structure of GFO and (d) distortion in oxygen polyhedra with cation-oxygen bond lengths and angles in GFO unit cell.

Fig. 2. (Color online) Temperature dependent Raman spectra of GFO single crystal (a) at selected temperatures between 18 K to 450 K and in the frequency range (b) 275-550 cm$^{-1}$ and (c) 620-820 cm$^{-1}$. Each spectrum was fitted with sum of Lorentzian line shapes and fitted spectra are superimposed.

Fig. 3. (Color online) Temperature dependence of the line shape parameters of nine modes between 250-550 cm$^{-1}$. The solid and dotted lines represent the simulated peak position and line width data according to eq. (3) and (4), respectively. The ferri to paramagnetic transition temperature ($T_c$ = 290 K) is marked for reference.

Fig. 4. (Color online) Plot of $\omega(T)$ vs. $\left\{\dfrac{M(T)}{M_s}\right\}^2$ for mode M5. The lines are linear least square fits to the data and $T_f$ (~ 180 K) marks the point at which abrupt change of slope ($m$) occurs.

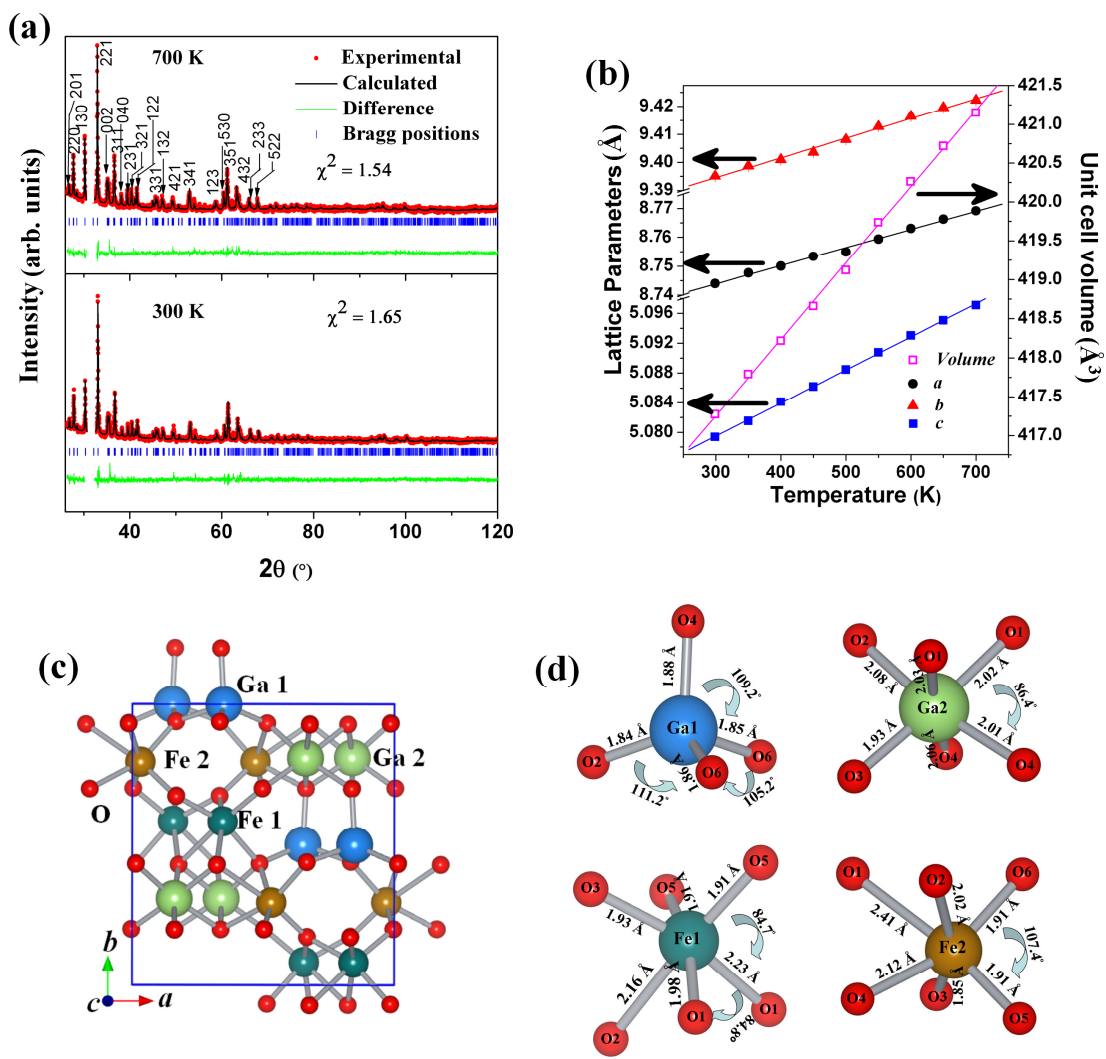

Fig. 1- Mukherjee *et al*.

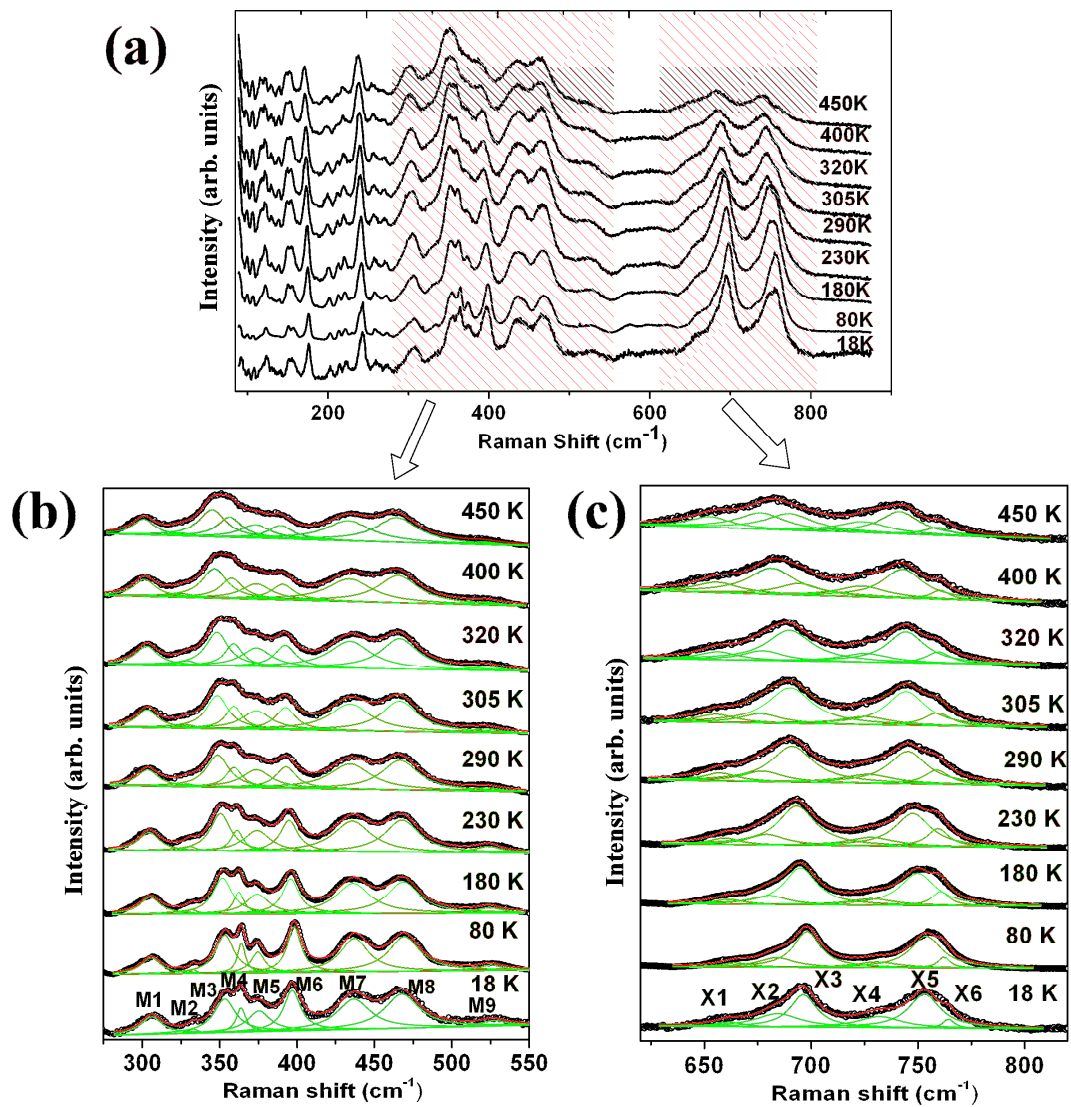

Fig. 2- Mukherjee *et al.*

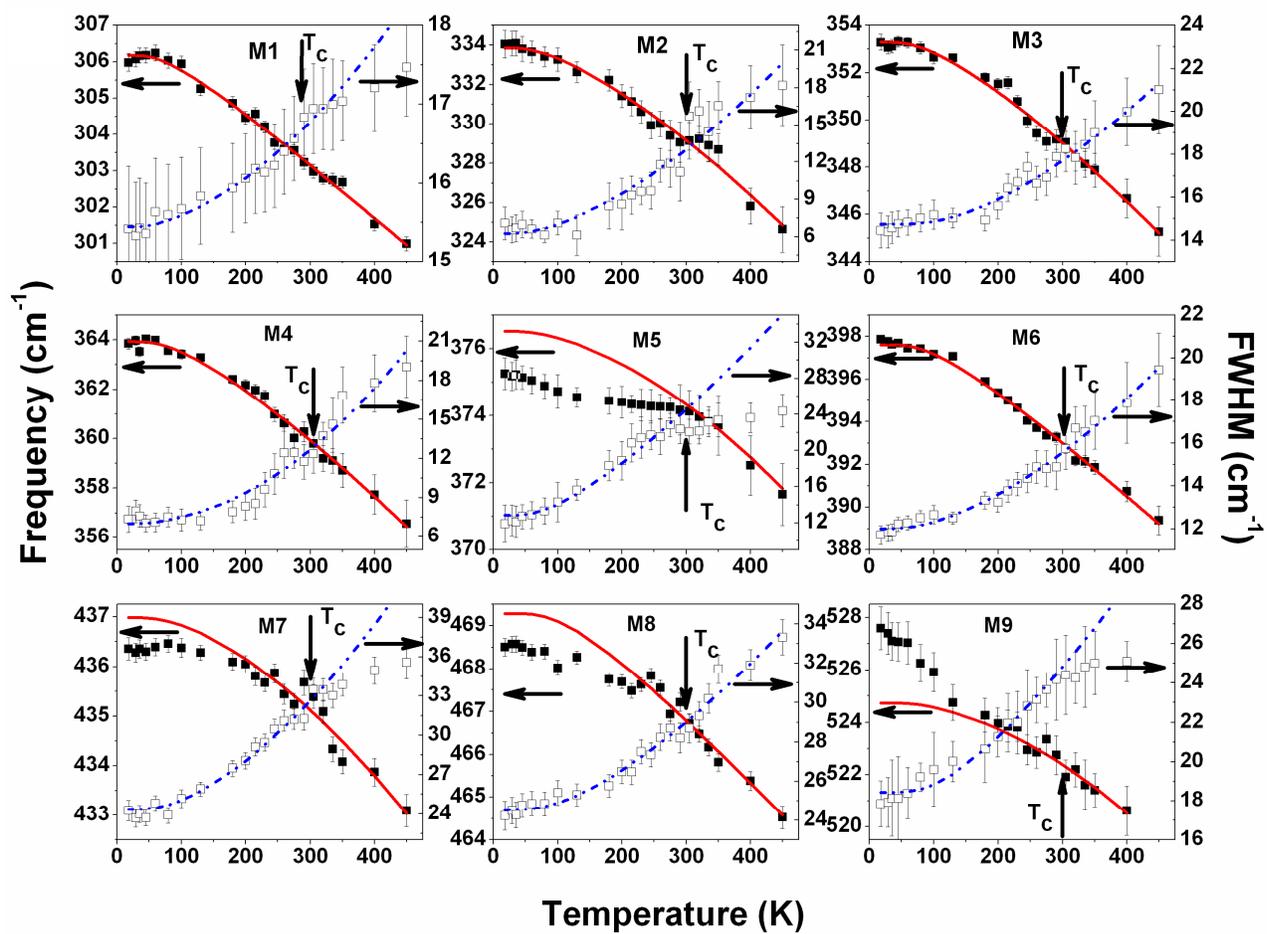

Fig. 3- Mukherjee *et al.*

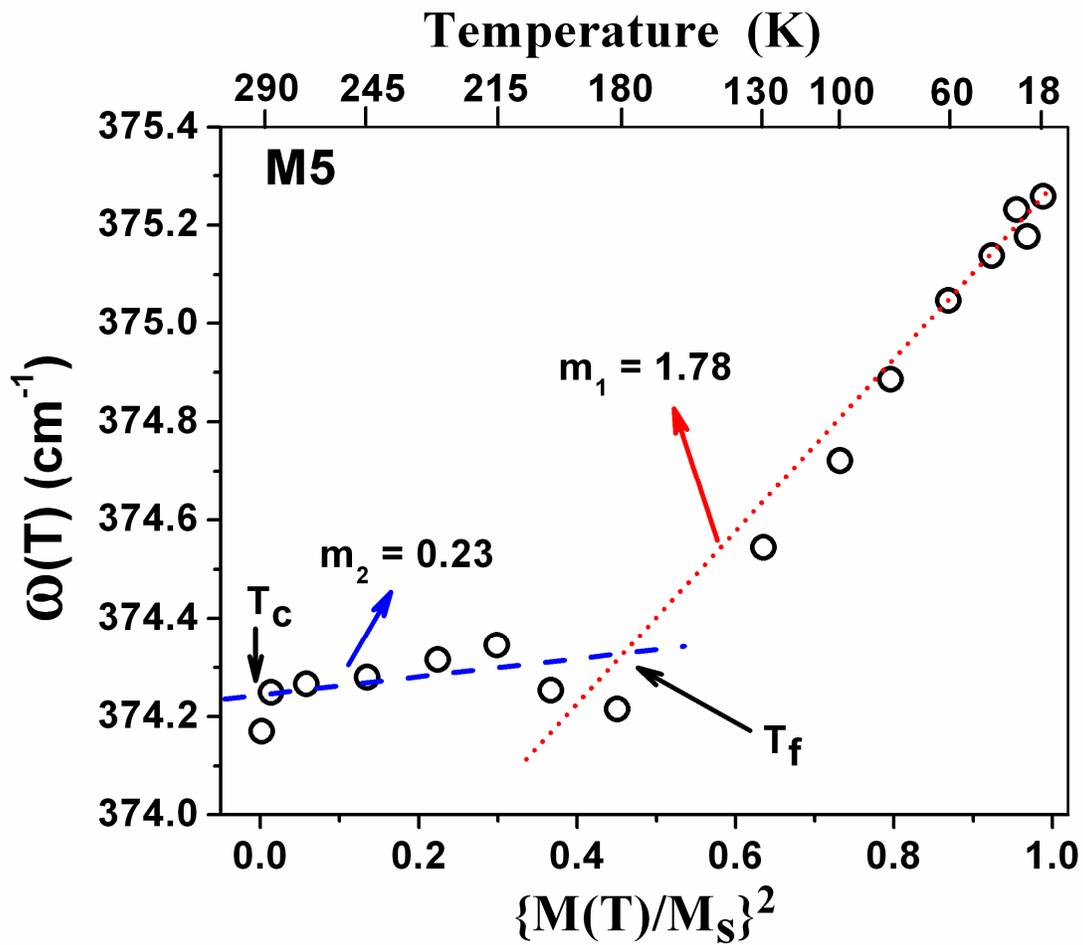

Fig. 4- Mukherjee *et al*.